\documentclass[12pt]{article}

\usepackage[mathscr]{eucal}
\usepackage{amssymb,latexsym}
\usepackage{verbatim}
\usepackage{graphicx}
\usepackage{amsmath}
\usepackage{amsthm}
\usepackage{enumerate}
\usepackage{authblk}
\usepackage{color}
\usepackage{url}

\usepackage{caption}

\usepackage[normalem]{ulem}

\newtheorem{theorem}{Theorem}[section]
\newtheorem{lemma}[theorem]{Lemma}
\newtheorem{conj}[theorem]{Conjecture}
\newtheorem{thm}{Theorem}[section]
\newtheorem{lem}[thm]{Lemma}
\newtheorem{Def}{Definition}[section]
\newtheorem{prop}[thm]{Proposition}

\DeclareMathOperator{\Ric}{Ric}

\newcommand\norm[1]{\left\lVert#1\right\rVert}

\title{Cosmological singularities from high matter density without global topological assumptions}

\author{Martin Lesourd}

\affil{University of Oxford, UK}

\begin{document}
\date{}
\maketitle
\vspace{.2in}

\begin{abstract}
Cosmological singularity theorems such as that of Hawking and Penrose assume local curvature conditions as well as global ones like the existence of a compact (achronal) slice. Here, we prove a new singularity theorem for chronological spacetimes that satisfy what we call a `past null focusing' condition. Such a condition forces all null geodesics \(\gamma:[0,a)\to M\) with future endpoint \(\gamma(0)\) to develop a pair of conjugate points if past complete. By the Einstein field equations, such a condition will be satisfied if the density of matter fields remains sufficiently high towards the past of the spacetime, as may be expected in certain cosmological scenarios. The theorem obtained doesn't make starting assumptions about the spacetime's topology, such as the existence of a compact achronal slice, and if in addition to a `past null focusing' condition we assume the timelike convergence condition, then further consequences pertaining to the existence of CMC foliations and the character of the singularity are obtained. With the addition of the timelike convergence condition, we obtain the conclusion that all timelike geodesics are past incomplete, rather than the existence of a single incomplete non-spacelike geodesic. 
\end{abstract}



\section{Introduction}

\hspace*{0.2in} Many cosmological singularity theorems are based on certain global topological assumptions. Hawking's original singularity theorem \cite{H} is a prime example since it assumes the existence compact spacelike slice with mean curvature having a strict sign. A more recent example is the result of Galloway and Ling \cite{GL} which demonstrates (under the null energy condition) the null incompleteness of spacetimes admitting `expanding' compact Cauchy surfaces that aren't topologically spherical spaces\footnote{A three dimensional Riemannian manifold is said to be a spherical space if its topology is $S^3\backslash \Gamma $ where $\Gamma$ is a subgroup of $SO(4)$}. \\ \indent 
The goal of this paper is to present a new cosmological singularity theorem that doesn't involve starting assumptions about the spacetime's topology. The absence of such an assumption is replaced by a quasi-local assumption on the Ricci tensor of the spacetime metric. We call this assumption a past null focusing condition because from the geometric point of view it has the effect of causing past inextendible null geodesics to develop conjugate points. Described in more detail below, the physical meaning behind this assumption is that the energy-density due to the presence of matter fields remains sufficiently strong in the past direction. \\ \indent   
If in addition to the past null focusing condition, we assume the timelike convergence condition (the geometric version of the strong energy condition), then the theorem leads to further consequences which are rather atypical from the point of view of standard cosmological singularity theorems. In particular, one can obtain the conclusion that \textit{all} timelike geodesics are past incomplete, which is much stronger than the usual scenario where one infers the existence of a \textit{single} incomplete geodesic.

\subsection{Context and Theorem}
\hspace*{0.2in} There is a classic so-called splitting theorem in Riemannian geometry due to Cheeger and Gromoll which demonstrates the rigidity associated with lines in the presence of non-negative Ricci curvature ; see chapter 9 of \cite{Petersen}.
\begin{thm}
Let \((V,h)\) be a complete Riemannian manifold with non-negative Ricci curvature. If \((V,h)\) admits a line then \((V,h)=(\mathbb{R}\times \Sigma,dt^2+m)\) where \(m\) is the Riemannian metric induced on \(\Sigma\) by \(h\).
\end{thm}
In this statement, a `line' is an inextendible curve every closed segment of which realizes the Riemannian distance between the segment endpoints. Yau conjectured that there hold a Lorentzian analog to the above theorem; see problem 115 of \cite{Y}. Less than a decade later and through the work of a number of authors, the following splitting theorem was formulated.
\begin{thm}[Lorentzian splitting theorem] 
Let \((M,g)\) be a spacetime that satisfies the timelike convergence condition, i.e., \(\Ric(u,u)\geq0\) for all timelike \(u\). Suppose that \((M,g)\) is either globally hyperbolic or timelike geodesically complete. Then, if \((M,g)\) contains a complete timelike line, \((M,g)\) splits as a globally static metric product \((\mathbb{R}\times \Sigma,-dt^2+h)\). 
\end{thm}
In this statement, a `timelike line' is an inextendible timelike curve every closed segment of which realizes the Lorentzian distance between the segment endpoints. \\ \indent 
The Lorentzian splitting theorem captures to some extent the rigidity of timelike geodesic completeness. In the spirit of Yau's conjecture, a stronger rigidity statement was conjectured by Bartnik \cite{Bart} for cosmological spacetimes, i.e., spacetimes with a compact Cauchy surface that satisfy the timelike convergence condition.
\begin{conj}[Bartnik \cite{Bart}]
A timelike geodesically complete cosmological spacetime must be a static, complete metric product \((\mathbb{R}\times \Sigma,-dt^2+h)\).
\end{conj}
By `static' we mean that the spacetime admits a global hypersurface orthogonal timelike Killing vector field.\\ \indent 
Bartnik's conjecture has been settled under a number of different auxiliary hypothesis, with the weakest to date due to Galloway and Vega \cite{GV}. \\ \indent 
Let us note here that Bartnik's conjecture admits an equivalent formulation in terms of null rays. Recall that a null ray is a future (or past) inextendible achronal causal curve with a past (or future) endpoint, and that a null line is an inextendible achronal causal curve without endpoint. The statement is then that a timelike geoedesically complete cosmological spacetime splits as in Bartnik's conjecture if and only if it does not admit a null ray.  \\ \indent 
That no null rays implies splitting is immediate for it is a standard consequence of Lorentzian geometry that a spacetime with compact Cauchy surface admits either a timelike or null line.\footnote{In fact, the admission of such a line can be deduced from much weaker assumptions, as in theorem 1.3 below; see for instance chapter 8 of \cite{BEE}.} Thus, if there are no null rays, then there are no null lines and so the line is timelike. If the spacetime is timelike geodesically complete, the timelike line is complete, and splitting follows from the Lorentzian splitting theorem. \\ \indent 
That a cosmological spacetime which splits as in the Bartnik conjecture has no null rays can be seen as follows.\footnote{The following argument is simpler than our original argument and was pointed out to us by anonymous referee.} If \((M,g)\) splits as in Bartnik's conjecture then \((M,g)=(\mathbb{R}\times \Sigma, -dt^2+h)\) and the spacetime is static. Thus there exists a complete timelike Killing vector field \(k^a\). By flowing along the integral curves of \(k^a\), we can define a projection map \(f:M\to \Sigma\) which takes curves in \(M\) and projects them into \(\Sigma\). Such a map, it can be shown, takes a Lorentzian ray and turns it into Riemannian ray. And so in particular for some null ray \(\eta\subset M\) with endpoint \(p_0\in \Sigma_0\), we have \(f:\eta\to \delta \subset \Sigma_0\) where \(\delta\) is a Riemannian ray in \(\Sigma_0\) with endpoint \(p_0\). But now \(\Sigma_0\) admits a ray, which it cannot by compactness. The crux of the argument is that staticity allows for a projection map which turns a Lorentzian ray into a Riemannian ray. This property of the projection map was studied and justified in Appendix B of \cite{Gallsub}, albeit in a slightly different context, but we do not reproduce those arguments since we shall not use them in what follows. \\ \indent 
The above remarks suggests that null rays are related to splitting. Thus we can expect that conditions prohibiting the existence of null rays lead to strong consequences. To study this, we shall consider the effects of imposing any local geometric inequality on \(\Ric_g(n,n)\) for \(n\) null which under some circumstances would prohibit the occurrence of null rays. More precisely, we shall consider the following definition. 
\begin{Def}
A spacetime is past null focusing if \(Ric_g(n,n)\) is such that, along every null geodesic \(\gamma\) with tangent \(n\), \(\gamma\) has a pair of conjugate points if past complete.
\end{Def} 
Such conditions have been considered in \cite{Tipler}, \cite{Minguzzi}, \cite{Borde}. An explicit example, which here we express in the future direction, is the following; see Prop. 1 of \cite{Tipler}. For all future complete (affinely parametrized) null geodesics \(\gamma:[0,\infty)\to M\), require that \[\lim_{s\to \infty}  s\int_s^{\infty} \Ric_g(n,n)ds' >1 \] where \(n\) is the tangent vector to \(\gamma\) at \(\gamma(s)\). Note that by the Einstein field equations this translates to an assumption about the growth of the energy density of matter fields towards the future. In the past direction, the assumption will apply to cosmological models in which the matter density remains sufficiently high towards the past.\\ \indent 
The point of such past null focusing conditions is that they rule out the possibility of past null rays if the spacetime is past null complete. This is because past null complete geodesics are forced to develop conjugate points and thus fail to remain achronal. The main result of this article may now be stated below. It is atypical in that other known cosmological singularity theorems with low causality conditions invariably make global topological assumptions (eg., the existence of a compact achronal slice). In the theorem below these assumptions are bypassed in favor of the extra geometric condition of past null focusing. Another somewhat unusual feature is the result (b)(ii) which guarantees that every timelike geodesics is past incomplete, which is significantly stronger than the usual conclusion that there exists a single non-spacelike incomplete geodesic. 
\begin{thm} 
Let \((M,g)\) be a chronological spacetime that satisfies a past null focusing condition. Then there are two cases:
\begin{enumerate}
\item[(1)] \(M\) is past null incomplete, 
\item[(2)] \(M\) is past null complete, in which case, we have the following:
\begin{enumerate}
\item[(a)] \(M\) is globally hyperbolic, there is a single TIF, and the Cauchy surfaces of \(M\) are compact with finite fundamental group.
\item[(b)] If in addition \(M\) satisfies the timelike convergence condition, then:
\begin{enumerate} 
\item[(i)]  \(M\) admits a compact CMC Cauchy surface \(S\) such that the past \(J^-(S)\) is foliated by compact CMC Cauchy surfaces,
\item[(ii)] \(M\) is either a complete static metric product \[(M,g)=(\mathbb{R}\times \Sigma, -dt^2+h)\] or admits an incomplete timelike line, and moreover if this line is past incomplete, then every timelike geodesic is past incomplete.
\end{enumerate}
\end{enumerate}
\end{enumerate}
\end{thm}
By a `CMC' Cauchy surface \(S\subset M\) we mean a Cauchy surface \(S\) with mean curvature \(H_{S}=tr_{h}(K)\) constant on \(S\) where we are writing the constaint equations of the Einstein field equations as \[R_h+{(tr_h(K))}^2-\norm{K}_h^2=\rho\] where \(R_h\) is the scalar curvature associated with the initial data set \((S,h,K)\), \(h\) is the Riemannian metric induced on \(S\) by \(g\), \(K\) is the second fundamental form of \(S\) in \(M\), \(\rho=T(u,u)\) where \(u\) is the timelike vector field normal to \(S\) and \(T(\cdot,\cdot)\) is the stress energy tensor appearing in the Einstein field equations. The interest in inferring the possibility of a CMC foliation stems from their various useful properties. See \cite{HD} for a recent introductory review of their significance and a list of open questions.\\ \indent
By geometrization and the Poincar\'e conjecture, the restriction to a finite fundamental group means that the slice is a spherical space (i.e., topologically \(S^3\backslash \Gamma\) where \(\Gamma\) is a subgroup of \(SO(4)\)). \\ \indent
We also observe that for chronological spacetimes satisfying a past null focusing condition, the above theorem to deduce the existence of null incompleteness if any one of the conditions within (a) and (b) is violated; for instance, if the fundamental group is infinite, or, under the timelike convergence condition, if there lacks a CMC Cauchy surface or if the spacetime is not static.  \\ \indent 
Consider also the following theorem of Minguzzi, which he proves as a means of capturing a possible yet rarely considered cosmological scenario.
\begin{thm}[Minguzzi \cite{Minguzzi}]
Let \((M,g)\) be a spacetime of dimension greater than \(2\). If \((M,g)\) is null geodesically complete, chronological, contains a future trapped surface, satisfies the timelike convergence condition, the generic condition, together with a past null focusing condition, then \((M,g)\) is globally hyperbolic with compact Cauchy surface \(S\) and has an incomplete timelike line. 
\end{thm}
The proof of this result relies on a result of Minguzzi that chronology along with the absence of either past (future) null rays leads to their being a single TIF (TIP) and the existence of a compact Cauchy surface. Here, we note that the current argument will also imply that the Cauchy surface must have finite fundamental group, and that there will be a foliation by CMC Cauchy surfaces towards the past. \\ \indent 
Finally, we note that theorem 1.3 is likely to admit a generalization based on an averaged form of the timelike convergence condition. This depends on whether it is possible to generalize the Lorentzian splitting theorem with such a condition. We plan to address this in a further work. Such a version would be preferable since there are mounting doubts as to the physical suitability of the timelike convergence condition in the context of inflation and early cosmology.

\section{Preliminaries}
\hspace*{0.2in} Our conventions are basically as in \cite{BEE}. A spacetime \((M,g)\) is an \(n\)-dimensional Lorentzian manifold. A spacetime \((M,g)\) is said to satisfy chronology (causality) if there are no closed timelike (causal) curves in \(M\). An open set \(U\subset M\) is said to be causally convex is no non-spacelike curve intersects \(U\) in a disconnected set. Given a point \(p\in M\), the spacetime \((M,g)\) is said to be strongly causal at \(p\) if \(p\) has arbitrarily small causally convex neighborhoods. The spacetime is strongly causal if it is strongly causal at every point. \\ \indent A spacetime \((M,g)\) is globally hyperbolic if and only if it is causal and \(J^+(p)\cap J^-(q)\) is compact for all \(p,q\in M\). Standard causality theory (see chapter 3 of \cite{BEE}) shows that this is equivalent to the existence of a Cauchy surface \(S\), i.e., so that \(M=D(S)\) where \(D(\cdot)\) denotes the domain of dependence defined in terms of causal curves. \\ \indent We recall the following definition of an edge; as in chapter 14 of \cite{BEE}.
\begin{Def}
Let \(S \subset M\) be achronal. Then \(p \in S\) is an edge point of \(S\) provided every neighborhood \(U(p)\) of \(p\) contains a timelike curve \(\gamma\) from \(I^-(p,U)\) to \(I^+(p,U)\) that does not meet \(S\). We denote by \(edge(S)\) the set of edge points of \(S\).
\end{Def}
From this follows the following standard result of causality theory. 
\begin{prop}
Let \(S\) be closed. Then each \(p \in \partial I^+(S)\backslash S\) lies on a null geodesic contained in \(\partial I^+(S)\backslash S\), which either has a past endpoint on \(S\), or else is past inextendible in \(M\).
\end{prop}
A spacetime \((M,g)\) is said to be causally disconnected by a compact set \(K\subset M\) if there exists two infinite sequences of points \(\{p_i\},\{q_i\}\) with \(q_i \leq p_i\), which diverge to infinity, such that for any \(i\), all future directed non-spacelike curves from \(p_i\) to \(q_i\) intersect \(K\). Here , an infinite sequence in a non-compact topological space is said to `diverge to infinity' if given any compact subset \(C\), only finitely many elements of the sequence are contained in \(C\).\\ \indent 
The Lorentzian distance function \(d(p,q)\) is defined as in chapter 4 of \cite{BEE}. We note the following useful lemma, which appears as corollary 4.7 in \cite{BEE}.
\begin{lem}
For any globally hyperbolic spacetime \((M,g)\), the Lorentzian distance function is continuous and \(d(p,q)\) for \(p,q\in M\) is finite.
\end{lem}
We shall also briefly touch on the notion of the causal boundary of a spacetime; see chapter 6 of \cite{BEE} for an introduction. The key notions we shall use stem from the work of \cite{GKP}. The causal boundary of a spacetime was developed with the aim of describing some of the spacetime's global causal and geometric properties, by attaching to it a notion of a boundary representing the `edge' of the spacetime. The important sets in this construction are terminally indecomposable past or future sets, i.e., TIP or TIF, which are constructed as follows. \\ \indent 
A past (future) set \(A\) is a necessarily open set in \(M\) such that \(I^{-(+)}(A)=A\). An indecomposable past (future) set is a set that cannot be written as a union of two proper subsets both of which are past (future) sets. A terminally indecomposable past (future) set is an indecomposable past (future) set which is not the chronological past (future) of any point in the spacetime. We note the following key lemma which will be used below. 
\begin{lem}[Geroch-Kronheimer-Penrose \cite{GKP}]
A subset \(W\) of the strongly causal spacetime \((M,g)\) is a TIP (TIF) if and only if there exists a future (past) directed inextendible timelike curve \(\gamma\) such that \(W=I^{-(+)}(\gamma)\). 
\end{lem}
In the arguments to follow, we shall need a number of results from \cite{GL}. We refer our reader to \cite{GL} for the relevant definitions, to avoid repeating these here.

\section{Proof}

\hspace*{0.2in} First we show the statement in (a). Suppose past null geodesic completeness. In that case, the past null focusing condition forces \((M,g)\) to have no past null rays. Given that \((M,g)\) is chronological, the following theorem of Minguzzi \cite{Minguzzi} implies that \((M,g)\) is globally hyperbolic with a single TIF.
\begin{thm}[Minguzzi \cite{Minguzzi}]
If \((M,g)\) is chronological and without past null rays then \((M,g)\) is globally hyperbolic and has a single TIF, given by \(M\).
\end{thm}
Owing to an argument in Penrose's \cite{Penrose} singularity theorem, we can further argue that the Cauchy surface must be compact. \\ \indent In particular, let \(\Sigma\) be a Cauchy surface for \((M,g)\). Let \(V\subset \Sigma\) be any compact spacelike submanifold with codimension \(2\) with non-empty edge. By standard causality theory, \(\partial I^-(V)\) is a set which is generated by achronal null geodesics which are either future inextendible or with future endpoint intersecting \(edge(V)\). Given that \((M,g)\) is null geodesically complete and globally hyperbolic, the former possibility does not occur. Thus, all such generators intersect \(edge(V)\). By the absence of past null rays, it cannot occur that these achronal null geodesics are past inextendible. Thus, the null hypersurface \(\partial I^-(V)\) must end. Since \(\partial I^-(V)\) is closed, it is also compact. By the standard homeomorphism constructed in the proof of the Penrose singularity theorem, eg., see \cite{Penrose} or chapter 9 of \cite{Wald}, it follows that \(\Sigma\) must be compact. \\ \indent 
So \((M,g)\) now admits a compact Cauchy surface, and has a single TIF as described in (a). It remains to show that \(S\) has finite fundamental group. \\ \indent 
Suppose otherwise that \(S\) has infinite fundamental group. Then \(S\) has a non-compact Riemannian universal cover \(\tilde{S}\). In that case \(\tilde{S}\) is a non-compact Cauchy surface for a spacetime \((\tilde{M},\tilde{g})\) which is the Lorentzian universal covering spacetime for \((M,g)\). The following lemma, described in Galloway and Ling \cite{GL2}, justifies this fact. 
\begin{lemma}[Galloway, Ling \cite{GL2}]
Let \(V\) be a smooth spacelike Cauchy surface in a spacetime \((M,g)\) having induced metric \(h\) and second fundamental form \(k\). Suppose \(\phi:\tilde{V}\to V\) is a Riemannian covering, with metric \(\tilde{h}=\phi^* h\) on \(\tilde{V}\). Then there exists a Lorentzian covering \(\Phi : \tilde{M}\to M\), with metric \(\tilde{g}=\Phi^* g\) on \(\tilde{M}\), such that \(\tilde{V}\) is a Cauchy surface for \((\tilde{M},\tilde{g})\) with induced metric \(\tilde{h}\) and second fundamental form \(\tilde{k}=\Phi^*k\).
\end{lemma}
Recall that a past null focusing condition is a condition along all null geodesics \(\gamma\subset M\) and that it is imposed on \(\Ric_g(n,n)\) where \(n\) is tangent to \(\gamma\). We note then that \((\tilde{M},\tilde{g})\) also satisfies a past null focusing condition. In particular, the covering map \(\Phi:\tilde{M}\to M\) lifts any null geodesic \(\gamma\subset M\) to a null geodesic \(\tilde{\gamma}\), i.e., \(\gamma=\Phi(\tilde{\gamma})\), with the relevant contraction of the Ricci tensor of \(\tilde{g}\) given by \(\Phi^*(\Ric_g(n,n))=\Ric_{\tilde{g}}(\tilde{n},\tilde{n})\) with \(\tilde{n}\) now tangent to \(\tilde{\gamma}=\Phi^{-1}(\gamma)\).  \\ \indent 
Given that past null completeness of \((M,g)\) implies past null completeness of \((\tilde{M},\tilde{g})\), we have that \((\tilde{M},\tilde{g})\) is past complete, and thus that \((\tilde{M},\tilde{g})\) has no past null rays. But then by the arguments above \((\tilde{M},\tilde{g})\) has a compact Cauchy surface, which is a contradiction. \\ \\ \indent
It is worth noting here that although the past null focusing condition as defined lifts to the covering spacetime constructed in lemma 3.1, the absence of null rays \textit{per se} is not a condition that lifts. Take for instance a spatially identified version of \(4\)-dimensional Minkowski spacetime \((\mathbb{R}\times \mathbb{T}^3,\eta)\), this spacetime has no null rays due to the failure of achronality, and yet there is an abundance of null rays in its covering spacetime \((\mathbb{R}^4,\eta)\). \\ \\ \indent 
We now assume the timelike convergence condition so as to get the statements in (b)(i). If \((M,g)\) is timelike geodesically complete, the absence of past null rays then implies that \((M,g)\) has a complete timelike line and thus splits as in the Bartnik conjecture. \((M,g)\) is then a static, complete metric product. \\ \indent 
So \((M,g)\) is either past null geodesically complete and timelike geodesically incomplete, or past null incomplete. Note that in the former case, the spacetime is either past timelike geodesically incomplete, or future, or both, and that by the compactness of its Cauchy surface, it must admit an incomplete timelike line. \\ \indent
Now consider the statement (b)(ii). The main conclusion here follows essentially from work of Tipler \cite{Tipler}. Past null completeness and the conditions of theorem 1.3 implies the absence of past null rays. From Minguzzi's theorem 3.1, the past causal boundary \(C^-\) consists of a single element \(c\) and the spacetime is globally hyperbolic. Then we use the following result of \cite{Tipler}.
\begin{thm}[Tipler \cite{Tipler}]
If a non-flat stably causal spacetime \((M,g)\) satisfies the timelike convergence condition, with equality holding only if \(\Ric_g=0\), and \((M,g)\) has a past causal boundary \(C^-\) consisting of a single point, then there exists a point \(p \in M\) such that through \(p\) there passes a \(C^{2,\alpha}\) Cauchy surface \(S\) with constant mean curvature, and furthermore \(I^-(S)\) can be uniquely foliated by \(C^{2,a}\) Cauchy surfaces with constant mean curvature.
\end{thm}
Note that Tipler's assumption concerning the equality case of the timelike convergence condition is a stronger overall condition than the timelike convergence condition. As described in \cite{Tipler} however, this extra component is only used for the uniqueness part of his statement. For the existence of a CMC Cauchy surface, his argument relies only on the standard timelike convergence condition. \\ \indent
We now show that every timelike geodesic is past incomplete. This will follow from the following proposition, applied in the time reversed direction. 
\begin{prop}
Let \((M,g)\) be a chronological spacetime with a single TIP. Suppose that \((M,g)\) admits a future incomplete timelike ray \(\gamma:[0,a)\to M\), i.e., a timelike curve with past endpoint \(\gamma(0)\) every segment of which has Lorentzian length which realizes the Lorentzian distance between the segment endpoints. Then all timelike geodesics are future incomplete. 
\end{prop}
Note below that this proposition may be given a quicker proof than the one offered below; we nevertheless include the argument below since the construction is explicit and may be considered in more general circumstances.\footnote{We gratefully acknowledge this to an anonymous referee.} This shorter argument is done by using directly the condition that there is a single TIP and the presence of the incomplete ray \(\gamma\) and it may be sketched as follows. Let \(\gamma\) be the future incomplete timelike ray, and let \(\delta:[0,b)\to M\) be any future-inextendible timelike curve, not necessarily a geodesic. Because \(M\) has a single TIP, it follows that \(\delta \subset I^-(\gamma) = I^-(\delta)\). Let \(t_1 \in [0,b)\) such that \(\gamma(0) \subset I^-(\delta(t_1))\). It then follows, using the TIP and ray conditions, that \(L(\delta_{[t_1,b)})<L(\gamma)\), and hence that \(L(\delta)< L(\delta_{[0,t_1]}) + L(\gamma)\).
\begin{proof}
As above, chronology and a single TIP means that there are no future null rays and that the spacetime admits a compact Cauchy surface. \\ \indent 
Now let \(\gamma:[0,a)\to M\) be an affinely parametrized future incomplete timelike ray, and let \(\delta:[0,b)\to M\) be any affinely parametrized future inextendible timelike geodesic. Since both are future inextendible timelike curves, lemma 2.2 implies that each of defines its own TIP. Since there is only a single TIP, we must have \(I^-(\gamma)= I^-(\delta)\). We'll show that \(\delta\) must be future incomplete. To do so, we shall repeatedly use the standard fact that for a globally hyperbolic spacetime, the Lorentzian distance function \(d(p,q)\) is finite and continuous, and furthermore, that for any two causally related points \(p\leq q\), there always exists a causal curve, necessarily a geodesic, that realizes the Lorentzian distance between the points; see chapter 3 and 4 of \cite{BEE} for a review of the properties implied by global hyperbolicity. \\ \indent 
Consider two sequences \(\{p_i\}\) and \(\{q_i\}\) of points lying on, respectively, \(\gamma\) and \(\delta\) with \(i\in \mathbb{N}\), \(p_1=\gamma(0)\) and \(q_1=\delta(0)\). Choose the sequence such that \(p_i\) and \(q_i\) lie, respectively, in the chronological past of \(p_{i+1}\) and \(q_{i+1}\). By belonging to the same TIP, we may also choose this sequence to be such that for each triple \(p_i,q_i,p_{i+1}\), there exists three future directed causal curves connecting these points as follows. \(\eta_i\) connects \(p_i\) to \(p_{i+1}\), \(\alpha_i\) connects \(p_i\) to \(q_{i}\), and \(\beta_i\) connects \(q_i\) to \(p_{i+1}\). We may choose these points and these curves so that each curve realizes the Lorentzian distance, which is either zero in the case of points connected by an achronal causal curve or strictly positive otherwise. In virtue of \(\gamma\) being a timelike ray, and \(\{p_i\}\) being points on \(\gamma\), we can take the curves \(\eta_i\) to be segments of \(\gamma\).  \\ \indent 
By the reverse triangle inequality, we have \[d(p_i,p_{i+1}) \geq d(p_i,q_i)+d(q_i,p_{i+1}) \] and \(d(p_i,p_{i+1})>0\). By considering \(p_i\) further along \(\gamma\), we must have that \(d(p_i,p_{i+1}) \to 0\) as \(i\to \infty\) and moreover that \(d(p_1,p_i)<c\) for all \(i\) where \(c\) is a positive constant. Since the Lorentzian distance function is always non-negative, it follows that both \(d(p_i,q_i)\) and \(d(q_i,p_{i+1})\) approach \(0\) as \(i\to \infty\).\\ \indent 
Now consider the point \(p_1\), which is the first in the sequence \(\{p_i\}\). Using once more the reverse triangle inequality, we have \[d(p_1,p_{i+2}) \geq d(p_1,q_{i+1})+d(q_{i+1},p_{i+2}) \] From above we have \(d(q_{i+1},p_{i+2}) \to 0\) as \(i\to \infty\) and \(d(p_1,p_{i+1}) <c\). Thus for all \(i\) we have \[d(p_1,q_{i+1})< c\] We now show that the boundedness of \(d(p_1,q_{i+1})\) for all \(i\) leads to a contradiction. \\ \indent 
Since \(\delta\) is future complete, its length \(l(\delta)\) is infinite, and thus \(d(q_1,q_i) \to \infty\) as \(i\to \infty\). Using once more the reverse triangle inequality we have \[d(p_1,q_i)\geq d(p_1,q_1)+d(q_1,q_i)\] which now implies that \(d(p_1,q_i)\to \infty\) as \(i\to \infty\), which contradicts the previous statement that \(d(p_1,q_{i+1})<c\) for all \(i\). Thus \(\delta\) must be future incomplete.
\end{proof} 
So if \((M,g)\) contains a past incomplete timelike line, and thus a past incomplete timelike ray, then in fact every timelike geodesic is past incomplete.\\ \\ \indent 
Finally, we note that the proposition is false if the future incomplete geodesic \(\gamma\) is not a future timelike ray.\footnote{We gratefully acknowledge that this was pointed out to us by an anonymous referee.} By spatially identifying an example originally considered by Busemann and Beem \cite{BB}, one can construct a globally hyperbolic spacetime with a single TIP containing future incomplete geodesics as well as future complete timelike geodesics. See in particular chapter 3 of \cite{BEE} for a discussion of their example.

\end{document}